\documentclass[prx,nobibnotes,nofootinbib,twocolumn,showpacs]{revtex4-2}

\usepackage{mathrsfs}

\usepackage{calc}
\usepackage[T1]{fontenc}
\usepackage[latin1]{inputenc}
\usepackage{amsmath}
\usepackage{amssymb}
\usepackage{xspace}
\usepackage{graphicx}
\usepackage[normalem]{ulem}
\allowdisplaybreaks[1]

\newcommand{\be}{\begin{equation}}
\newcommand{\ee}{\end{equation}}

\usepackage{xcolor}

\definecolor{darkgreen}{rgb}{0.2,0.6,0}
\definecolor{lightblue}{rgb}{0,0.5,0.8}
\definecolor{lightred}{rgb}{0.8,0.2,0.2}
\definecolor{darkorange}{rgb}{1,0.549,0}
\definecolor{brown}{rgb}{0.609, 0.164, 0.164}

\makeatother

\begin{document}

\definecolor{rvwvcq}{rgb}{0.08235294117647059,0.396078431372549,0.7529411764705882}
\definecolor{wrwrwr}{rgb}{0.3803921568627451,0.3803921568627451,0.3803921568627451}

\title{Regular black holes with stable cores}
\author{Alfio Bonanno}
\affiliation{INAF, Osservatorio Astrofisico di Catania, via S.Sofia 78 and INFN, Sezione di Catania, via S. Sofia 64, I-95123,Catania, Italy}
\author{Amir-Pouyan Khosravi}
\author{Frank Saueressig}
	\affiliation{Institute for Mathematics, Astrophysics and Particle Physics (IMAPP), \\ Radboud University, Heyendaalseweg 135, 6525 AJ Nijmegen, The Netherlands}

\begin{abstract}
Non-singular black hole geometries typically come with two spacetime horizons: an (outer) event horizon and an (inner) Cauchy horizon. This nurtures the speculation that they may be subject to a mass-inflation effect which renders the Cauchy horizon unstable. We analyze the dynamics associated with spherically symmetric, regular black holes taking the full backreaction between the infalling matter and geometry into account. On this basis, we identify the crucial features taming the growth of the mass function and diminishing the curvature singularity at the Cauchy horizon. It is demonstrated explicitly that the regular black hole solutions proposed by Hayward and obtained from Asymptotic Safety satisfy these properties.
\end{abstract} 
\maketitle

\section{Introduction}
The detection of gravitational waves emitted from the merger of two black holes \citep{Abbott:2016blz,Abbott:2016nmj,Abbott:2017vtc} and the first picture of an event horizon published by the event horizon telescope \citep{Akiyama:2019cqa} have moved black holes into the focus of observational science. This raises the intriguing question whether the black holes observed in nature are indeed the black holes known from general relativity or black hole ``mimickers'' resembling these objects. From the viewpoint of classical general relativity the theoretical description of black holes is very simple: a Schwarzschild black hole includes a spacetime singularity, creating the curvature of spacetime, which is hidden behind an event horizon. The latter realizes the cosmic censorship hypothesis \citep{he}, stating that spacetime singularities can not be observed by exterior observers. Moreover, the no-hair theorem \cite{Israel:1967wq,Israel:1967za,Carter:1971zc} states that black holes in general relativity are characterized completely by their mass, charge, and angular momentum. Any perturbation created, e.g., during the formation of a black hole in a stellar collapse, dies off quickly, ensuring that the configuration settles to the simple, stationary solution. 

The occurrence of singularities is often taken as an indicator that general relativity is incomplete and should be generalized to a quantum theory expelling this feature. In anticipation of such a theory, various {\it ad hoc} modifications of the Schwarzschild solution have been proposed \citep{bardeen1968non,Borde:1996df,dy92,Dymnikova:2001fb,hayward,Frolov:2017rjz}, also see \citep{fr14,ansoldi} for a reviews. Commonly, these modifications respect the limiting curvature hypothesis \citep{1982ZhPmR..36..214M,Markov:1984ii,Polchinski:1989ae,Frolov16} and replace the black hole singularity by a regular piece of de Sitter space. As a consequence, the (outer) event horizon is supplemented by an (inner) Cauchy horizon. The consistency of these models is challenged by the observation that the Cauchy horizon is generally unstable to external perturbations \citep{penrose68}. In particular, the self-consistency arguments forwarded in \citep{cr18} suggest that regular black hole geometries are not viable models for describing the black holes observed in nature since the effect of mass inflation renders the Cauchy horizon exponentially unstable against perturbations. In this work, we perform the first complete dynamical analysis of this instability. 
In conclusion, we show that there are specific regular black hole solutions where the divergence of the mass function is suppressed before the Cauchy horizon is reached at very large advanced times. While this weakens the singularity at the Cauchy horizon, the resulting geometries violate strong cosmic censorship and may therefore not be globally hyperbolic. In this situation, it becomes crucial to also account for the energy loss due to Hawking radiation which may radically change the interior geometry.	

\section{Setup and General Framework}
A generic spherically symmetric line element can always be written as
$ds^2=g_{ab} dx^a dx^b + r^2 d\Omega^2$, $a,b=0,1$,  
where  $r=r(x^a)$  is the radius of the 2-spheres with $x^a$ being constant.  It is convenient to introduce a generalized Schwarzschild mass function $M(x^a)$  by means of the gradient of $r(x^a)$,\footnote{In the sequel, we set Newton's constant $G=1$, implying that all dimensionful quantities are measured in Planck units.}  
\begin{equation}\label{Mdef}
g^{ab}\partial_a r \partial_b r = f(x^a) = 1 - \frac{2 M(x^a)}{r} \, .
\end{equation}
The taxonomy of different static black hole geometries is encoded in the radial behavior of $M(r)$. For a Schwarzschild black hole $M = m$ where $m$ is the mass of the object at large distances $r\gg m$.   
In the physical picture proposed by Sakharov \citep{sakharov},  a phase transition to a false vacuum occurs
at Planckian distances from the center so that a de Sitter core eventually develops \citep{dy92,frolov98} and  $M(r)\sim r^3$ for small $r$. An explicit model realizing such a phase transition is the Hayward model \citep{hayward} where
\begin{equation}
\label{hw}
M(r) = \frac{m r^3}{r^3 +2 m l^2} \, . 
\end{equation}
Here $l$ is a free parameter whose value should be fixed by the underlying quantum gravity model. Other suggestions for regular black hole geometries include the Bardeen black holes \citep{bardeen1968non,AyonBeato:2000zs}, 
the asymptotically safe black holes  \citep{bonanno2000,Koch:2014cqa}, Planck stars \citep{Rovelli:2014cta,DeLorenzo:2014pta,Saueressig:2015xua}, and the loop black hole
described in \cite{loop}, also see \citep{Frolov16} for additional references and discussion.

\begin{figure}
\centering
\includegraphics[width=0.5\textwidth]{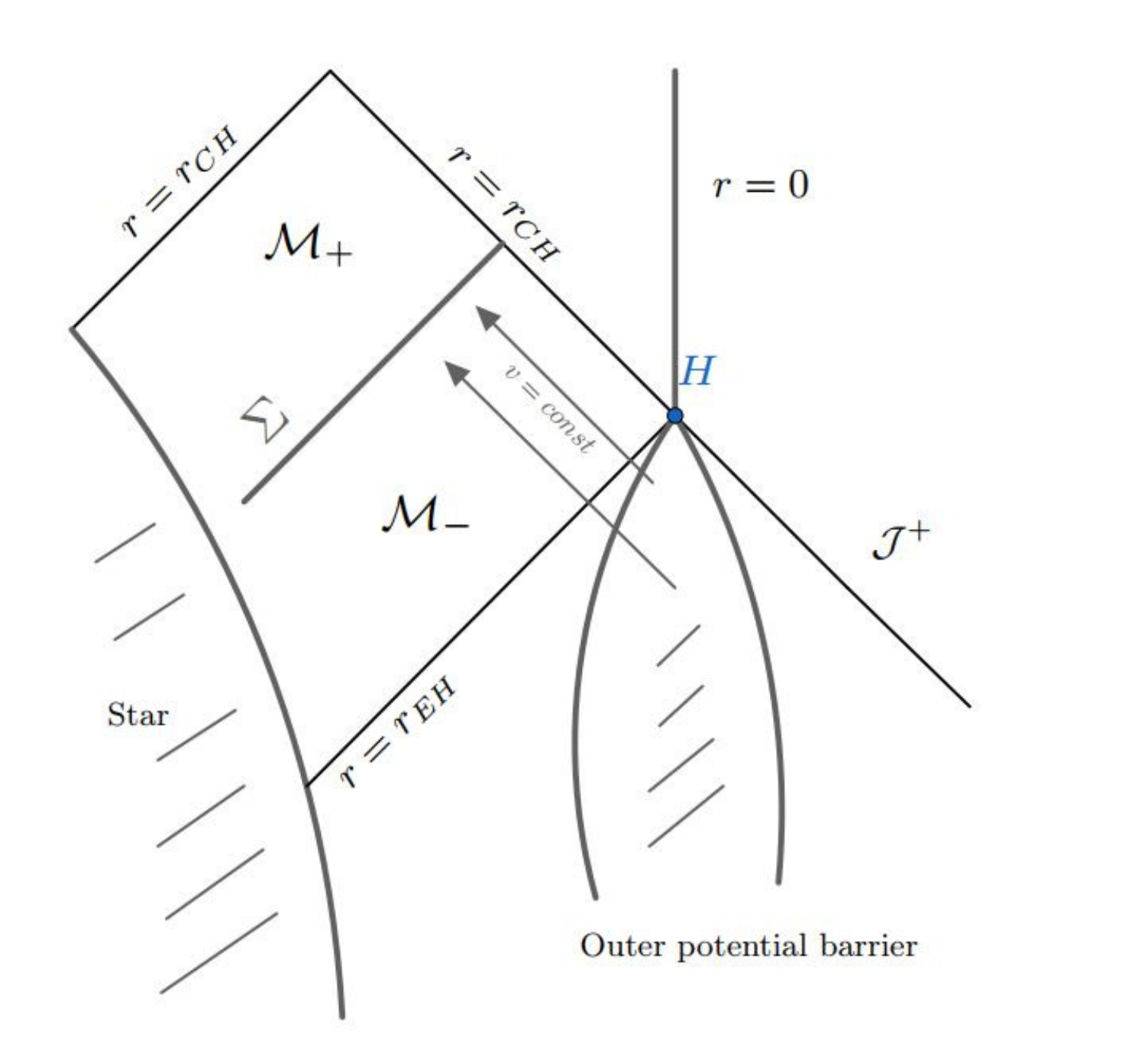}
	\caption{The Ori model of the interior of a regular black hole. 
	The $r=0$ hypersurface is not 
singular in these models.
Radiation enters the black hole backscattered from the outer potential barrier
as ingoing null rays.
Outgoing radiation is schematically represented by a thin shell 
$\Sigma$ which crosses the Cauchy horizon. The ``corner'' $H$ is a singular point of  the conformal 
diagram. At variance with the classical Ori model, no singularity develops 
in the future sector of the shell.\label{ori}}
\end{figure}
The occurrence of a de Sitter core induces a non-trivial topology change  in the black hole geometry. In variance with the Schwarzschild case, the horizon condition $f(r) = 0$ now has two solutions: beside the event horizon (EH) located at $r_+$ there is also an inner, Cauchy horizon (CH) at a smaller radius $r_{-} < r_{+}$. For the Hayward model \eqref{hw} these are given by the two positive roots of
\be\label{eq:hayhor}
2 l^2 m -2 m r^2+r^3=0 \, . 
\ee
As a consequence of the modified horizon structure, the spacetime  is no longer globally hyperbolic. We define the surface gravity $\kappa_\pm$ at the horizons by
\be\label{surfacegrav}
\kappa_\pm \equiv \pm \, \frac{1}{2} \, \left. \frac{\partial f(r)}{\partial r} \right|_{r = r_\pm} \, , 
\ee 
where the sign ensures $\kappa_\pm > 0$. The event horizon emits Hawking radiation with temperature $T_{\rm EH} = \kappa_+/(2 \pi)$. Owing to the resulting energy loss, the coordinate distance between the event and Cauchy horizon decreases. A complete evaporation is attained in an infinite time, as measured by an stationary observer at asymptotic infinity. The final, asymptotic configuration is a regular, extremal black hole with vanishing surface gravity. The mass $m_{\rm cr}$ of the remnant follows form the condition that the position of $r_\pm$ coincide. For the Hayward model,
$
m_{\rm cr} = \frac{3\sqrt{3}}{4} l
$.

 A classical analysis reveals that the Cauchy horizon  is generally unstable to external perturbations \citep{penrose68}. For this reason there have been doubts about the consistency of regular black hole models \citep{fr17}.  The physical picture underlying the analysis is as follows: the collapse of a mass distribution to form a black hole will lead to the emission of a stream of gravitational waves when the black hole settles into its final ``hairless'' state. A part of this wave tail will be reflected by the gravitational potential at $r > r_+$, creating an ingoing flux of positive energy crossing the event horizon. In the vicinity of the Cauchy horizon a part of this flux will again be back-scattered by the gravitational potential in the black hole interior, creating an outgoing positive energy flux. 
A consistent calculation which  takes into account the combined effect of the ingoing radation from the collapsing star and the backscattered gravitational radiation near the CH shows that a  genuine scalar singularity develops at the Cauchy horizon \cite{poisson89,*poisson90}. This situation is illustrated in Fig.\ \ref{ori}.

Technically, it is convenient to formulate this analysis in terms of the coordinates $(v,r)$ where time is parameterized by the ingoing Eddington-Finkelstein coordinate $v$, defined by the condition that radially ingoing light rays follow curves with $v = \text{const}$. One then introduces a  time-dependent perturbation in the mass function $M(r) \rightarrow M(r,v)$ 
describing a shell of outgoing 
lightlike dust on the interior geometry of the black hole. The corresponding energy flux is encoded in the Isaacson  effective energy-momentum tensor \cite{isaacson68a,*isaacson68b} and its decay follows Price's law \citep{price72a,price72b}. The behavior of  the curvature invariants near the Cauchy horizon is rather insensitive to the details of the local fields trapped inside the event horizon. The rate of divergence of  the Coulomb component of the Weyl curvature\footnote{Here $C_{\mu\nu\rho\sigma}$ denotes the Weyl tensor and $\{l^\mu,n^\mu,m^\mu, \bar{m}^\mu \}$ is a complex null-tetraed with $l^\mu n_\mu = -1$, $m^\mu \bar{m}_\mu = 1$. $\Psi_2$ is the only non-zero scalar for a Petrov type-D spacetime.} \cite{Novikov:1989sz},  $\Psi_2 \equiv C_{\mu\nu\rho\sigma} l^\mu m^\nu \bar{m}^\rho n^\sigma $, can then be characterized by  the \textit{anomalous dimension} of the instability
\begin{equation}
\label{anomalous}
\nu = \frac{d \ln |\Psi_2| }{d \ln v} \, . 
\end{equation}
In the limit $v \rightarrow \infty$, $\Psi_2 \simeq c_1 \, v^{-p} \,  {\rm e}^{\kappa_- \, v}$ with $c_1$ being a $v$-independent constant, entailing that the anomalous dimension behaves as $\nu \simeq \kappa_{-} \, v - p$.\footnote{Here and in the following we use the $\simeq$-symbol to denote the asymptotic behavior of a quantity in the limit $v \rightarrow \infty$.} Since the divergence in $\Psi_2$ is triggered by the exponential increase of $M(r,v)$, this effect has has been dubbed {\it mass-inflation}.
It is difficult to imagine an efficient self-regulator mechanism at work before the curvature has curbed the core at  Planckian levels \citep{anderson93,bore99}:  at variance with Petrov type-N, type-D curvatures cannot
be confined in thin layers, and it is reasonable to expect that the complete past evolution of the 3-dimensional 
hyper-cylinder $r=r_{-}$ is singular.

This conclusion is less compelling, however, if the backreaction of the Hawking radiation in the interior is taken into account. In this case, the Cauchy horizon also receives a
blue shifted influx of negative energy density originating from  
the event horizon. This flux could
 halt the contraction of the generators of the Cauchy horizon well before the curvature has 
reached Planckian levels, at least for mini black holes \citep{bal91}.  Drawing a definite conclusion therefore requires comparing 
the characteristic time scales of the evaporation process and the growth time of the
mass-inflation instability.

Such an analysis has recently been presented in \cite{cr18}, describing the evaporation process of the regular black hole   
through a  sequence of {\it quasi-adiabatic} states for which 
\begin{equation}
\label{stability}
\nu_E \equiv  -\frac{d \ln \kappa_+}{d \ln v} \ll \nu \, ,
\end{equation} 
always. 
This led to the conclusion that the characteristic time scale associated with the evaporation process is  too small to efficiently act as a  self-regulator for the interior geometry. 

The estimation of the exponent $\nu$, controlling the time scale on which the instability builds up, is however more delicate than in the original Poisson-Israel model because of the non-linear functional form of the quasi-local mass function $M(r,v)$ in these models. 
A simple self-consistent estimation of the mass dynamics based on the DTR relations \cite{dray,*redmount} 
might therefore not be sufficient to consistently describe 
 the contraction of the generators spanning the Cauchy horizon and 
the energy influx near the ``corner'' region $H$ in Fig.\ \ref{ori}.
In fact, the Cauchy horizon singularity of the classical  Poisson-Israel model is rather weak: 
the metric coefficient are still regular and the integrated tidal forces are bounded.
This necessitates a  
consistent calculation of the back-reaction of the radiation on the Cauchy horizon of a regular black hole, taking both the positive energy influx due to classical perturbations and the backreaction of the regular geometry into account. This work presents the results of this analysis, describing the dynamics of a 
 realistic gravitational collapse based on the Ori model \citep{ori91} of the classical Poisson-Israel model \cite{Israel:1967wq,Israel:1967za}. 

  \begin{figure*}[!hbt]
      \includegraphics[width=0.48\textwidth]{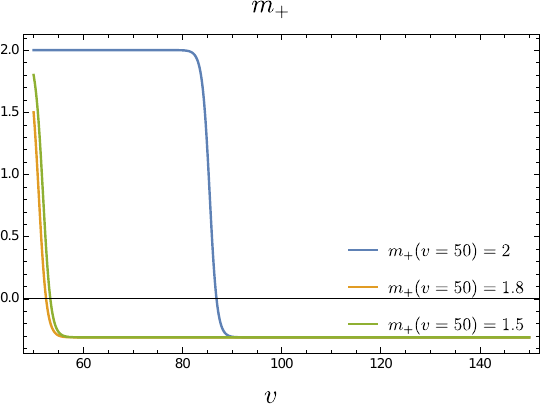}
      \quad
      \includegraphics[width=0.48\textwidth]{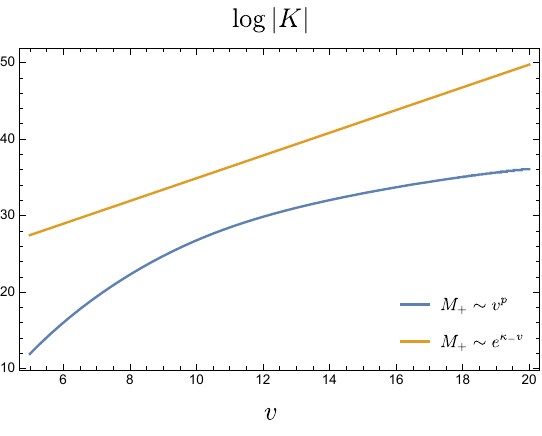}
      \caption{Left: the dynamical evolution of $m_{+}$ in the future sector of the 
      	shell for various initial conditions and $l=1/2$ and $p=11$. In the past sector of the shell $m_0=2$. For $v \rightarrow \infty$ the solution settles on its asymptotic value $-r_-^3/(2l^2)$. 
      	Right: the asymptotic behavior of the Kretschmann scalar 
      	$K=R_{\alpha\beta\gamma\delta} R^{\alpha\beta\gamma\delta}$ in logarithmic
      	scale, for $l=1/2$, $m_0=2$ and $p=11$. Note that the
      	limiting behavior is insensitive to the initial condition for $m_{+}$.}
      \label{fig:evol}
  \end{figure*}
In the Ori model  the outgoing energy flux is modelled 
by a thin pressureless null shell $\Sigma$ which divides the spacetime in two regions  
${\cal M}_\pm$,  inside $(+)$ and outside ($-$) the shell. This shell acts as a catalyst to trigger the divergence of the mass function. 
Assuming spherical symmetry, the metric in each sector of spacetime can be written as
\begin{equation} \label{metric}
ds^{2} = -f_{\pm} (r,v_{\pm}) dv_{\pm}^2+2 dr dv_{\pm} + r^{2} d\Omega^2 \, , 
\end{equation}
where $f_{\pm}=1 - 2 M_\pm(r,v_{\pm})/r$.
The equality of the induced metric on $\Sigma$
forces $r$ to be the same on ${\cal M}_\pm$. For this reason it is convenient to choose $r$ as a parameter (not necessarily affine) along the null generators $s_{\pm}^\mu = d x_{\pm}^\mu/dr = (2/f_{\pm}, 1, 0, 0)$ of $\Sigma$. Since the shell moves light-like, we have 
\be\label{geoshell}
f_- dv_- = 2 dr \, .  
\ee 
Einstein's equations on each sector of the spacetime can then
be expressed in terms of the mass function $M(r,v)$, 
\begin{equation}
\frac{\partial M}{\partial r} = -4\pi r^2 T_v{}^v \, , \qquad
\frac{\partial M}{\partial v} = 4\pi r^2 T_v{}^r \, ,  
\end{equation}
together with $T_{rr} = 0$.
Continuity of the flux across $\Sigma$ requires \citep{bar91} 
\begin{equation}
\label{cont}
    [T_{\mu\nu} s^\mu s^\nu]=0 \, , 
\end{equation}
where the square brackets indicate the ``jump'' of a scalar quantity across the shell. This condition is consistent with the assumption of $\Sigma$ being pressureless. In terms of the lapse and mass-function, eq.\ \eqref{cont} then reads
\be\label{cont2}
\left. \frac{1}{f_+^2} \frac{\partial M_+}{\partial v_+} \right|_{\Sigma}= \left. \frac{1}{f_-^2} \frac{\partial M_-}{\partial v_-} \right|_{\Sigma} \, ,
\ee
with the derivatives evaluated before substituting the position of the shell. In general $v_{+}$ and $v_{-}$ have to be distinguished because $v_+$ is not just the continuation of $v_-$ to $\mathcal{M}_+$. The functional dependence of the two  is fixed by noting that the position of the null hypersurface $\Sigma$ in the two coordinate systems is
\begin{equation}
\label{rad}
f_{+} dv_{+} = f_{-} dv_{-} \;  \quad {\rm along} \; \Sigma \, . 
\end{equation}In the following we shall use (\ref{rad})  to express all physical quantities in terms of $v\equiv v_{-}$. Eq.\ \eqref{cont2} can then be recast as
\be\label{eq:mpmaster}
\left. \frac{1}{f_+} \frac{\partial M_+}{\partial v} \right|_{\Sigma} = F(v)
\ee
where
\be\label{defF}
F(v) \equiv \left. \frac{1}{f_-} \frac{\partial M_-}{\partial v} \right|_{\Sigma} \, . 
\ee
Denoting the position of the shell by $R(v)$ and derivatives with respect to $v$ by a dot, eq.\  \eqref{geoshell} provides a geodesic equation encoding the motion of the shell as a function of $v$
\be\label{eq:r2}
\dot{R}(v)= \left. \frac{1}{2} f_- \right|_{\Sigma} \, .
\ee
This relation allows to eliminate $f_-$ from \eqref{defF}, yielding
\be\label{defF2}
F(v) \equiv \left. \frac{1}{2 \dot{R}} \, \frac{\partial M_-}{\partial v} \right|_{\Sigma} \, .
\ee

\section{Mass-inflation effect in regular black holes}
The framework developed in the previous section allows to study the mass-inflation effect for a wide class of modified black hole solutions. The present discussion will focus on static, non-extremal black holes and neglects effects related to the black hole evaporation process.\footnote{The effect of Hawking radiation will be discussed elsewhere \cite{Bonanno-inprep}.} We first exemplify the analysis for the case of the Hayward geometry \eqref{hw}. The analysis of other regular black holes follows the same lines and we summarize our results in Table \ref{Tab.2}.  

\subsection{The Hayward geometry}
The Hayward geometry corresponds the mass function $M(r,v)$ given in (\ref{hw}), entailing
\be
f_\pm = 1- \frac{2 m_\pm(v_\pm) r^2}{r^3 +2 m_\pm(v_\pm) l^2} \, . 
\ee
Substituting $f_-$ into \eqref{eq:r2} gives a geodesic equation determining the position of the shell 
\be\label{r2}
\dot{R}(v)= \frac{1}{2}-\frac{m_{-} \, R^2}{R^3+2 \, l^2\, m_{-}} \, ,
\ee
where it is understood that $R$ and $m_\pm$ are $v$-dependent functions specifying the position of $\Sigma$ and the mass density. Eq.\ \eqref{eq:mpmaster} furthermore gives a differential equation for the mass function $m_+$
\be
\label{r1}
\frac{R^6 \, \dot{m}_{+}}{(R^3 + 2 l^2 m_{+}) (R^3-2 m_{+}(R^2-l^2))} = F(v) 
\ee 
where the explicit form of $F(v)$ reads
\be\label{eq:Feval}
F(v) = \frac{R^6 \, \dot{m}_{-}}{(R^3 + 2 \, l^2 \, m_{-}) (R^3-2 m_{-}(R^2-l^2))} \, . 
\ee
The system \eqref{r2} and \eqref{r1} is a coupled system of non-linear differential equations encoding the evolution of $R(v)$ and $m_+(v)$. 

We are now going to discuss the dynamics entailed by the system in detail. The boundary condition at the event horizon is determined by the Price's tail behavior so that in the past sector of the shell the mass function $m_-(v)$ can be written as
\begin{equation}\label{pricetail}
m_{-}(v) = m_0 - \frac{\beta}{(v/v_0)^{p}} 
\end{equation}
Here $m_0$ is the black hole mass and $\beta>0$ is a quantity with the dimension of a mass and $v_0$ is the initial value of $v$ which is set to one in the sequel. Since the dynamics of the shell is independent of $m_+(v)$, we can first study the motion of the shell based on \eqref{r2}. Since the horizons $r_\pm$ are, by definition, zeros of $f$, they constitute fixed points for $R(v)$. Placing the shell between the event and the Cauchy horizon, it will move inward and settle at $r_-$ for asymptotically large time. The position of the shell for asymptotically large values $v$ can be determined analytically by applying the Frobenius method. Substituting the asymptotic expansion
\be
R(v) = r_- + \frac{1}{v^s} \sum_{k=0}^\infty \frac{a_k}{v^k} \, , 
\ee
with $s>0$ and $a_0$ non-zero by construction, 
into eq.\ \eqref{r2} and equating powers of $v$ shows that a non-trivial solution requires $s=p$. The coefficients $a_k$ can be determined recursively, revealing that the asymptotics of $R(v)$ for large values $v$ (and $p > 2$) has to follow
\be\label{eq:Rvasym}
	R(v)-r_- \simeq \frac{r_- \, \beta \, v^{-p}}{4 \, m_0^2 \, \kappa_- }
	\left(1+ \frac{p}{\kappa_- v} +\cdots \right ) \, .
\ee

Substituting eqs.\ \eqref{pricetail} and the asymptotics \eqref{eq:Rvasym} into eq.\ \eqref{eq:Feval} yields the asymptotic behavior of $F(v)$:
\be\label{eq:Fasym}
F(v) \simeq -\frac{r_- \, \kappa_-}{2} 
\left ( 1-\frac{p+1}{{\kappa_-} v} +\cdots \right ) \, . 
\ee
Thus the right-hand side of \eqref{r2} tends to a finite, negative limit $F_\infty < 0$ as $v\rightarrow\infty$. The asymptotic behavior of $m_+$ can then be deduced from the Frobenius method 
\begin{equation}\label{eq:mphay}
	m_+(v) \simeq - \frac{r_-^3}{2 l^2} \left( 1 + \frac{3 \, \beta \, v^{-p}}{4 \, m_0^2 \, \kappa_-} + \cdots \right) \, .  
\end{equation}
Most remarkably, $m_+$ approaches a \emph{constant, negative value} as $v \rightarrow \infty$. This is the key difference to the standard mass-inflation scenario \cite{ori91}, where $m_+(v)$ grows exponentially. In this sense, the Hayward geometry does not experience a mass-inflation instability. This feature can readily be traced back to the denominator appearing at the right-hand side of eq.\ \eqref{r1}: for the Reissner-Nordstr\"om solution considered in \cite{ori91} the denominator is linear in $m_+$ indicating that the equation is solved by exponential functions. In contrast to this, the Hayward geometry leads to a quadratic dependence in $m_+$. Asymptotically, the solution then approaches a root of this polynomial which ensures that $m_+$ remains bounded.   

The asymptotic behavior \eqref{eq:mphay} can be confirmed from \eqref{r1} directly by substituting the asymptotic values $R(v) \sim r_-$ and $F(v) \sim - \frac{r_- \kappa_-}{2}$ and solving the simplified equation analytically
\be
\begin{split}
m_+(v) \simeq & \, - \frac{r_-^3}{2 l^2} \, \frac{e^{\kappa_- v} - l^2 \, e^{2 r_-^5 c} }{e^{\kappa_- v} + e^{2 r_-^5 c} \left(r_-^2 - l^2\right)}
\end{split}
\ee
where $c$ is an integration constant. For $v \rightarrow \infty$ both the numerator and denominator grow exponentially and $m_+(v)$ remains finite. This asymptotic behavior is also readily confirmed by integrating eqs.\ \eqref{r2} and \eqref{r1} numerically for varying initial conditions, see the left panel of Fig.\ \ref{fig:evol} for explicit examples.

In order to exhibit the physical consequences of our findings, we first study the $v$-dependence of the mass function \eqref{hw} evaluated at $r_-$
\be\label{massasymptotic}
M_+(r_-,m_+(v)) \simeq \frac{2 \, r_-^3 \, \kappa_- \, m_0^2}{3 l^2 \beta} \, v^p + \cdots \, . 
\ee
Notably, $M_+$ diverges as $v \rightarrow \infty$ with the anomalous dimension \eqref{anomalous} being
\be\label{eq:pHay}
\nu \simeq p \, . 
\ee
In comparison to the standard mass-inflation scenario, this divergence is no longer exponential though. $M_+$ grows polynomially in $v$ only. This feature propagates into the curvature invariants constructed from the geometry. Evaluating the general expression for the Kretschmann scalar \eqref{eq:Kretschmann} for the Hayward geometry yields
\be
K = \frac{48 m^2 (32 l^8 m^4 - 16 l^6 m^3 r^3 + 72 l^4 m^2 r^6 - 8 l^2 m r^9 + 
	r^{12})}{(2 l^2 m + r^3)^6}
\ee
where $m \mapsto m_+(v)$ is the $v$-dependent mass function. Our interest is in the late-$v$ behavior of the curvature. This may readily be inferred by substituting the asymptotic expansion \eqref{eq:mphay}.  Notably, the leading term of this expansion cancels in the denominator. As a result
\be
K \simeq \frac{4}{9 l^4} \left( \frac{m_0^2 \, \kappa_- \, v^p}{\beta} \right)^6 \, 
\ee
is \emph{power-law divergent} as $v \rightarrow \infty$. Again, this is at variance with the standard mass-inflation scenario where $K$ grows exponentially in $v$. The two cases are compared in the right panel of Fig.\ \ref{fig:evol} 
showing the asymptotic $v$-dependence of $K$. We also observe that in our analysis the lapse function in the past sector of the shell approaches zero as a power law, at variance with what has been assumed in \citep{cr18} when discussing the DTR relations.

\subsection{Physics consequences of the modified mass function}
 A consequence of this modified behavior is that the singularity building at the Cauchy horizon becomes integrable. This can be seen as follows. Using the advanced coordinate in the future sector of the shell $v_+$, the asymptotic form of the metric near the Cauchy horizon reads
 \be\label{eq:cauchyasym}
 ds^2  \simeq 2 \frac{dv_+}{r} \left( r dr + m_+(v_+) dv_+ \right) + r^2 d\Omega^2 \, . 
 \ee
 The new coordinate $u$, defined through the relation $du = (rdr + m_+(v_+)dv_+)$, is 
 regular at the Cauchy horizon. The line-element \eqref{eq:cauchyasym} then becomes \cite{Bonanno:1994qh}
 \be
 ds^2 \simeq 2 \frac{dv_+ du}{r} + r^2 d\Omega^2 \, . 
 \ee
 This expression is manifestly regular at the Cauchy horizon. Since it is possible to find a coordinate system where the metric is regular, the singularity building up at the CH is rather weak. This fact has profound consequences: as already realized by Ori \cite{ori91}, and further investigated
 by Burko \cite{Burko:1999zv}, the mass-inflation singularity does not satisfy the necessary conditions to be strong in the Tipler sense \cite{Tipler:1977zza}.\footnote{According to Tipler, a null singularity is called ``strong'' if there exists at least one component
 of the Riemann tensor (in a parallelly propagated frame) which does not converge when integrated with respect to the affine parameter $\tau$ twice. The physical meaning of this requirement is that the tidal distortion is not finite as an observer crosses the singularity.}  A measure of the tidal distortion experienced by an observer is obtained by integrating the square of the Weyl curvature twice. In the case of the standard mass-inflation scenario one finds
 \be\label{eq:tidal}
 C_{\mu\nu\rho\sigma} C^{\mu\nu\rho\sigma} \propto \frac{1}{V^2 (\log(-V))^{2p}} \, , 
 \ee
 where $V \equiv -e^{-\kappa_- v}$ is a Kruskal coordinate adapted to the inner horizon and $V \propto \tau$ in this case. The tidal
 distortion is obtained by twice integrating \eqref{eq:tidal} and is therefore finite. It has further been argued by Ori
 that this behavior could be suffcient to determine a $C^1$ extension of the spacetime beyond the Cauchy horizon \cite{ori91}. However,
 according to Krolak, eq.\ \eqref{eq:tidal} still signals a strong singularity, as the expansion of the congruence is divergent \cite{Krolak:1987}: if the components of the Riemann tensor are integrated only once, the integral does not converge on the singularity. 
 
 In the case of the Hayward geometry, the divergence of the Weyl curvature is further weakened
 \be\label{eq:Hay3}
 C_{\mu\nu\rho\sigma} C^{\mu\nu\rho\sigma} \propto (\log(-V))^{6p} \, .
 \ee
 Notably, already the first integral of this quantity is finite at the singularity. At variance with the original Poisson-Israel model, the singularity developing in the Hayward geometry comes with a finite volume of the congruence \emph{and} expansion  so that it is a weak singularity in both the Tipler and the Krolak definition. We take this as strong evidence that the Hayward geometry admits a $C^1$ extension of the spacetime beyond the singularity. While the investigation of this point is beyond the scope of the present work this certainly warrants a more detailed analysis in the future.

\subsection{Final state of the regular geometry}
We now compare the result \eqref{eq:pHay} and the rate of evaporation $\nu_E$, cf.\ eq.\ \eqref{stability}.  The massloss due to Hawking radiation follows from the Stefan-Boltzmann law,
\begin{equation}
\label{sb}
-\frac{dm_{-}(v)}{dv} = \sigma_{\rm SB} \left ( \frac{\kappa_+(v)}{2\pi}\right)^4 \, A_{\rm EH}(v) \, , 
\end{equation}
where $ A_{\rm EH}$ and  $\kappa_+$ are the area and  the surface gravity of the event horizon.   Eq.\ (\ref{sb}) can be explicitly solved in the large $v$ limit,
\begin{equation}\label{asymptm}
m_-(v) = m_{\rm cr}+ const/v + O(v^{-2}) \, ,
\end{equation}
where $m_{\rm cr}$ is the mass of the asymptotic configuration. Substituting eq.\ \eqref{asymptm} into eq.\ \eqref{surfacegrav} shows that the outer  surface gravity $\kappa_+(v)  \sim v^{-1/2}$ so that the exponent controlling the evaporation time asymptotes to $\nu_{\rm E} \sim 1/2$. 
%
Therefore we conclude that for the Hayward geometry the  backreaction of the Hawking radiation acts on time-scales similar to the growth of the mass function. The final state of the geometry at $v\rightarrow \infty$ may then be given by a cold extremal black hole which is not destabilized by the mass-inflation effect. Notably, this asymptotic configuration is only reached after an infinite time-span, as measured by a static observer situated at asymptotic infinity. Whether the formation of the remnant entails the loss of information, as suggested in \cite{Giddings:1993vj}, or the information leaks out of the black hole during the evaporation process, as recently been suggested in \cite{Hooft:2016itl}, is currently open to debate and beyond the scope of our work.

\subsection{Other regular black hole geometries}
The asymptotic analysis of the previous subsections is readily generalized to other black hole geometries exhibiting a Cauchy horizon. Our results are summarized in Table 
\ref{Tab.2}, which also contains the Reissner-Nordstr\"om analysis for reference. The most remarkable feature revealed by this analysis is that regular black hole solutions may or may not suffer from the mass-inflation effect: depending on whether the denominator appearing on the right-hand side of eq.\ \eqref{r1} is linear or quadratic, the Misner-Sharp mass diverges either exponentially or as a power-law in $v$. Hence regular geometries of the Hayward and RG-improved type are safe from mass inflation while Bardeen-type geometries exhibit the same instability as the Reissner-Nordstr\"om black hole. This shows that there is no relation between the presence of a de Sitter core characterizing a regular geometry and the mass-inflation effect. It is the non-linear relation between the Misner-Sharp mass and $m_+$ which is the decisive feature. 
\begin{table*}[t!]
	\renewcommand{\baselinestretch}{1.2}
	\begin{tabular}{cccccc}
		\hline
		& $f(r)$ & $F_\infty$ & $m_+(v)$ & \quad $m_+$-dependence \quad & \quad mass-inflation \quad \\ \hline
		Reissner-Nordstr{\"o}m & $\Bigg.  1- \frac{2m}{r} + \frac{e^2}{r^2}$ & $- \frac{r_- \kappa_-}{2}$ & $\simeq c_1 e^{\kappa_- v} v^{-p}$ & linear & yes 
		\\ \hline
		Hayward solution \citep{hayward} & $\Bigg. 1 - \frac{2 m r^2}{r^3 +2 m l^2}$ & $- \frac{r_- \kappa_-}{2}$ & $\simeq - \frac{r_-^3}{2l^2}$ & quadratic & no \\[1.2ex]
		RG improved black holes \citep{bonanno2000} & \quad $\Bigg. 1 - \frac{2 m r^2}{r^3 +\omega( 2 r+ 9 m)}$ \quad & \quad $- \frac{r_- \kappa_-}{2}$ \quad & \quad $\simeq - \frac{r_-^3 + 2 r_-  \omega}{9 \omega}$ \quad & \quad quadratic \quad & no \\[1.2ex]
		Bardeen black hole \citep{bardeen1968non,AyonBeato:2000zs} & $\Bigg. 1 - \frac{2 m r^2}{(r^2+a^2)^{3/2}}$ & $- \frac{r_- \kappa_-}{2}$ & $\simeq c_1 e^{\kappa_- v}  v^{-p}$ & linear & yes \\[1.2ex] \hline
	\end{tabular}
	\caption{\label{Tab.2} Asymptotic values of the mass function $m_+(v)$ for various regular black hole geometries. For the Reissner-Nordstr\"om and Bardeen solutions, $m_+(v)$ grows exponentially while for the Hayward and renormalization group (RG)-improved geometries it remains finite. The quantities $l, \omega$, and $a$ appearing in the lapse function are free parameters which should be determined from quantum gravity.}
\end{table*}


\section{Outlook}

 The existence of regular black holes free from the mass-inflation effect warrants a more elaborate analysis about the final state which takes into account the effect of Hawking radiation at the full, dynamical level. Moreover, it would be interesting to see whether the regularization mechanism discovered in this work is also operative in other types of regular black hole solutions including the black holes discussed by Dymnikova \citep{dy92,Dymnikova:2001fb,alessia19}. 
  We hope to come back to these points in the future. 

\bigskip
\acknowledgments
\emph{Acknowledgements} 
We thank S.\ Liberati for fruitful communication on the initial version of the article. A.-P.\ K.\ thanks the Osservatorio Astrofisico di Catania for hospitality during the initial stages of this project. A.-P.\ K.\ acknowledges financial support by an ACRI Young Investigators Training Program. 
\medskip

\appendix
\section{Curvature invariants}
All geometries investigated in this work are of the form
\begin{equation}\label{metric2}
	ds^{2} = -f(r,v) dv^2+2 dr dv + r^{2} d\Omega^2 \, . 
\end{equation}
The lapse function $f(r,v)$ is related to the Misner-Sharp mass $M(r,v)$ by
\be
f(r,v) = 1 - \frac{2M(r,v)}{ r} \, . 
\ee
The Kretschmann scalar $K \equiv R_{\mu\nu\rho\sigma} R^{\mu\nu\rho\sigma}$ resulting from the metric \eqref{metric2} is conveniently expressed in terms of $M(r,v)$ and reads
\be\label{eq:Kretschmann}
\begin{split}
K = & \frac{48 M^2}{r^6} - \frac{64 M M^\prime}{r^5} + \frac{32 (M^\prime)^2}{r^4} \\ & \, + \frac{16 M M^{\prime\prime}}{r^4}  - \frac{16 M^\prime M^{\prime\prime}}{r^3} + \frac{4 (M^{\prime\prime})^2}{r^2} \, . 
\end{split}
\ee
Here the primes denote partial derivatives with respect to $r$ and we suppressed all arguments for the sake of readability. Since $K$ does not contain derivatives with respect to $v$, its $v$-dependence may be obtained by substituting $m \mapsto m(v)$.

%

\end{document}